\documentclass[sigconf,nonacm]{acmart}
\usepackage{rotating}
\usepackage{graphicx}
\usepackage{xcolor}
\usepackage{comment}
\graphicspath{ {./images/} }
\usepackage{multicol, blindtext}

\AtBeginDocument{%
  \providecommand\BibTeX{{%
    \normalfont B\kern-0.5em{\scshape i\kern-0.25em b}\kern-0.8em\TeX}}}

\setcopyright{acmcopyright}
\copyrightyear{2023}
\acmYear{2023}

\acmSubmissionID{123-A56-BU3}

\begin{document}

\title{Multimodal Search on Iconclass using Vision-Language Pre-Trained Models}


\author{Cristian Santini}
\authornote{All authors contributed equally to this research.}
\email{cristian.santini@fiz-karlsruhe.de}
\orcid{0000-0001-7363-6737}
\affiliation{%
  \institution{FIZ Karlsruhe}
  \country{Germany}
}
\affiliation{%
  \institution{Institute AIFB, Karlsruhe Institute of Technology}
  \country{Germany}
}
\author{Etienne Posthumus}
\authornotemark[1]
\email{etienne.posthumus@partners.fiz-karlsruhe.de}
\affiliation{%
  \institution{FIZ Karlsruhe}
  \country{Germany}
}
\author{Tabea Tietz}
\authornotemark[1]
\email{tabea.tietz@fiz-karlsruhe.de}
\affiliation{%
  \institution{FIZ Karlsruhe}
  \country{Germany}
}
\affiliation{%
  \institution{Institute AIFB, Karlsruhe Institute of Technology}
  \country{Germany}
}

\author{Mary Ann Tan}
\authornotemark[1]
\email{ann.tan@fiz-karlsruhe.de}
\affiliation{%
  \institution{FIZ Karlsruhe}
  \country{Germany}
}
\affiliation{%
  \institution{Institute AIFB, Karlsruhe Institute of Technology}
  \country{Germany}
}

\author{Oleksandra Bruns}
\authornotemark[1]
\email{oleksandra.bruns@fiz-karlsruhe.de}
\affiliation{%
  \institution{FIZ Karlsruhe}
  \country{Germany}
}
\affiliation{%
  \institution{Institute AIFB, Karlsruhe Institute of Technology}
  \country{Germany}
}

\author{Harald Sack}
\authornotemark[1]
\email{harald.sack@fiz-karlsruhe.de}
\orcid{0000-0001-7363-6737}
\affiliation{%
  \institution{FIZ Karlsruhe}
  \country{Germany}
}
\affiliation{%
  \institution{Institute AIFB, Karlsruhe Institute of Technology}
  \country{Germany}
}

\renewcommand{\shortauthors}{Santini et al.}

\begin{abstract}
 Terminology sources, such as controlled vocabularies, thesauri and classification systems, play a key role in digitizing cultural heritage. However, Information Retrieval (IR) systems that allow to query and explore these lexical resources often lack an adequate representation of the semantics behind the user's search, which can be conveyed through multiple expression modalities (e.g., images, keywords or textual descriptions). This paper presents the implementation of a new search engine for one of the most widely used iconography classification system, Iconclass. The novelty of this system is the use of a pre-trained vision-language model, namely CLIP, to retrieve and explore Iconclass concepts using visual or textual queries.
\end{abstract}



\maketitle

\section{Introduction}
Classification systems and other terminology sources, such as controlled vocabularies and thesauri, play a key role in the digital curation of cultural heritage objects. These lexical resources, are used by Galleries, Libraries, Archives and Museums (GLAM) to describe items in their collection; the Getty Vocabularies \footnote{https://www.getty.edu/research/tools/vocabularies/}~\cite{harpring_development_2010} or the Social History and Industrial Classification (SHIC) system\footnote{https://www.shcg.org.uk/About-SHIC} are some examples in the GLAM domain. Iconclass\footnote{www.iconclass.org}~\cite{waal_decimal_1968} is the \emph{de facto} standard for the classification of artistic representations and images. It is organized as a hierarchy which comprises 10 main categories, or notations, (abstract and non-representational art, religion and magic, etc.) and each category has several narrower terms as descendants in a tree-like structure. Notations in Iconclass are used to describe and index artworks based on their depicted objects (animals, deities, etc.) or artistic themes (biblical, mythological, etc.). Each notation is associated with a text which provides a label to the notation.

Currently, some limitations affect the accessibility of the digital platforms through which many controlled vocabularies, thesauri and classification systems are published. Specifically, the adoption of Information Retrieval (IR) techniques based on word occurrence, e.g. Term Frequency-Inverse Document Frequency (TF-IDF) or set theory, e.g. Boolean search, to retrieve entries in these lexical resources may create usability issues for non-expert users which are not familiar with the terminology of iconography in the Western arts. In order to solve this problem, word embeddings can be used to mitigate the aforementioned limitation, since they are able to encode word meanings into low-dimensional vectors, which alleviates the need for exact term matches. 

The main contribution of this work is the implementation of a multimodal search engine for Iconclass. This system is designed to leverage a vision-language model, i.e. OpenAI's CLIP~\cite{radford_learning_2021}, and a visual similarity search~\cite{posthumus_art_2022}, to retrieve Iconclass concepts, or notations, based on either text or image inputs from the user. Section~\ref{main} of this work discusses the adapted pre-trained Vision-Language model and the implemented search system, Section~\ref{eval} contains results from a small preference-based survey conducted among Iconclass users of varying art historical background and Section~\ref{conclusion} concludes the paper.

\section{Multimodal IR in Iconclass} \label{main}

Following the development of deep learning architectures trained on a single expression modality, which might consists either of texts, images, video, audio or graphs, researchers have recently investigated the potentials of training neural models on emerging patterns which can be expressed across multiple modalities, e.g. with both visual and textual features. CLIP (\emph{Contrastive Language-Image Pre-training})~\cite{radford_learning_2021} follows this intuition. Trained on a dataset of 400M \textit{(image-caption)} pairs collected from the internet, CLIP, given an image, must predict which out of a set of 32,768 randomly sampled text snippets it was paired with in the dataset. The main idea behind this model is to use two different encoders, one for images and one for texts, and by formulating the learning objective as a contrastive loss, the model is able to align text and images into a shared embedding space to learn diverse visual concepts. 

The new multimodal search system in Iconclass takes advantage of the aforementioned vision-language model to find appropriate Iconclass notations given a text or image as input. The system consists of a database of $\approx$500K images of artworks and photographs, hereinafter referred to as \emph{Images-DB}, and a Faiss index~\cite{johnson_billion-scale_2017} of 2GB which stores the embeddings of the images from the aforementioned database, retrieved from the pre-trained CLIP model of~\cite{radford_learning_2021}. \textit{Images-DB} collects 531,172 images described with Iconclass notations, which were kindly provided by the Arkyves initiative\footnote{www.arkyves.org}. 
The database contains 2,526,145 Iconclass notations, of which 90,347 were unique. 

\begin{figure}
    \centering \includegraphics[width=0.4\textwidth]{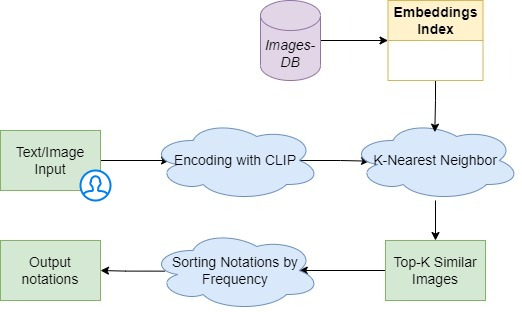}
    \caption{Diagram of the system architecture.}
    \label{fig:diagram}
\end{figure}

For visual search, the text or image initially given by the user as input is encoded into a multidimensional vector by using either the text or image encoder of CLIP. The obtained embedding is then used as seed for a similarity search on the Faiss index, which returns the top-K most similar annotated images in \emph{Images-DB}, by using a k-nearest neighbor algorithm. The output of the similarity search is then the set of similar images and the list of notations used to describe them, sorted by the number of assigned images in reversed order. A diagram of the underlying algorithm of the system is presented in Figure~\ref{fig:diagram}.

It is important to note that a user does not directly query the complete Iconclass hierarchy, but a set of annotated images. This represents a novelty since it indirectly exploits the information contained in the annotation of $\approx$500K samples in \emph{Images-DB}, which was carried by expert annotators. This multimodal search aims to exploit semantic similarity between text and images to retrieve, from a candidate set of images, potentially relevant notations provided by human experts, which is a relevant feature for users which are not well-versed in iconography. The new multimodal search engine was made available on a public demo\footnote{https://github.com/ISE-FIZKarlsruhe/iconclass/tree/main/multimodal}.

\section{Preference-based Evaluation} \label{eval}

Currently, Iconclass provides a TF-IDF based text search. Through a preference-based survey, the proposed multimodal search engine (\emph{System A}) was compared to the current approach (\emph{System B}).  For objectivity, the users were unaware of the system designation. 10 participants were gathered with a public call for volunteers.  Each respondent had to fill a spreadsheet containing overall 10 artwork images and 25 query strings. Given a query string, the top-10 results of system A and B were placed side-by-side. The users were then asked to select their preferred results, and to specify the reason for this preference: \emph{Preciseness}, the correctness of the returned notations (with respect to the image), and \emph{Exhaustiveness}, the recall of valid notations in the result list. Results from the survey are reported in Table~\ref{tab:comp}. 

Overall, respondents did not express a marked preference for one system over another. However, from the survey emerged that, when multimodal search was preferred, it was mainly due to exhaustiveness of the results. This result may derive from the fact that CLIP-based search does not aim to retrieve a single \emph{pinpoint} notation but a range of Iconclass codes related to a visual concept. For example, the query string \emph{Street} returns, when using multimodal IR, Iconclass notations which describe not only this iconographic element (\textit{25I141: street}), but also some related elements which are likely to occur in pictures of streets, such as humans (\textit{31D14: adult man}) or animals (\textit{34B11: dog}). The same does not happen for TF-IDF.

\begin{table}
\begin{center}
\begin{tabular}
{|p{0.3\linewidth}|p{0.3\linewidth}|p{0.25\linewidth}|} 
 \hline
 & Multimodal Search  & TF-IDF \\
 \hline
 \#Preferences & 105 & 104 \\ 
 \hline
 \#Preciseness & 64 & 72  \\ 
 \hline
 \#Exhaustiveness & 30 & 17  \\ 
 \hline
 \end{tabular}
 \caption{Results from preference-based survey aimed to compare visual-similarity search and TF-IDF search for Iconclass.
  \label{tab:comp}}
\end{center}
\end{table}

\section{Discussion and Conclusion} \label{conclusion}
This paper presents the new multimodal search engine of Iconclass. This system leverages pre-trained vision-language embeddings and a database of human-annotated images to return Iconclass notations based on either textual or visual inputs. The multimodal search engine offers users the advantage to query Iconclass with both images or free-text descriptions, which is a relevant feature to image curators which are not accustomed to the underlying vocabulary in Iconclass. However, the quality of CLIP-based results is still to be adequately estimated by using objective ground-truths. As a consequence, multimodal search can be exploited in order to complement search results from TF-IDF, rather than be considered as an equivalent alternative. As future work, the possibility to combine results coming from different search systems for Iconclass, e.g. both from visual similarity and word similarity, can and will be taken in consideration. Moreover, new features and services for Iconclass users will be introduced, such as the publication of a SPARQL endpoint to enable external web services to exploit the new similarity-based search.

\bibliographystyle{ACM-Reference-Format}
\bibliography{Iconclass+CLIP}

\end{document}